\documentclass[namedreferences]{SolarPhysics}
\usepackage[optionalrh]{spr-sola-addons} 
\usepackage{graphicx,color}        
\usepackage{color}           
\usepackage{url}             

\newcommand{\arcsec}{^{\prime\prime}}

\newcommand{\aap}{    {\it Astron. Astrophys.}}

\newcommand{\apj}{    {\it Astrophys. J.}}
\newcommand{\apjl}{   {\it Astrophys. J. Lett.}}
\newcommand{\apss}{   {\it Astrophys. Spa. Sci.}}

\newcommand{\jgr}{    {\it J. Geophys. Res.}}
\newcommand{\mnras}{  {\it Mon. Not. Roy. Astron. Soc.}}
\newcommand{\nat}{    {\it Nature}}

\newcommand{\pasj}{   {\it Pub. Astron. Soc. Japan}}

\newcommand{\solphys}{{\it Solar Phys.}}

\newcommand{\ssr}{    {\it Space Sci. Rev.}}

\begin{document}

\begin{article}

\begin{opening}

\title{{Slippage of  Jets  Explained by the  Magnetic Topology   
of NOAA Active Region 12035
}}

\author[addressref={aff1},corref,email={reetikajoshi.ntl@gmail.com}]{\inits{R.}\fnm{R.}~\lnm{Joshi}}
\author[addressref=aff2]{\inits{B.}\fnm{B.}~\lnm{Schmieder}}
\author[addressref=aff1]{\inits{R.}\fnm{R.}~\lnm{Chandra}}
\author[addressref=aff2]{\inits{G.}\fnm{G.}~\lnm{Aulanier}}
\author[addressref={aff2,aff3,aff4}]{\inits{F.P.}\fnm{F.P.}~\lnm{Zuccarello}}
\author[addressref=aff5]{\inits{W.}\fnm{W.}~\lnm{Uddin}}
\address[id=aff1]{Department of Physics, DSB Campus, Kumaun University, Nainital -- 263002, India}
\address[id=aff2]{Observatoire de Paris, LESIA, UMR8109 (CNRS), F-92195, Meudon Principal
Cedex, France}
\address[id=aff3]{Centre for mathematical Plasma Astrophysics, Department of Mathematics, KU Leuven, Celestijnenlaan 200B, B-3001 Leuven,
Belgium}
\address[id=aff4]{Mullard Space Science Laboratory, UCL, Holmbury St. Mary, Dorking, Surrey, RH5 6NT,UK}
\address[id=aff5]{Aryabhatta Research Institute of Observational Sciences (ARIES), Nainital -- 263002, India}

\runningauthor{R. Joshi {\it et al.}}
\runningtitle{Slippage of Jets Explained by the Magnetic Topology }

\begin{abstract}
In this study, we present the investigation of eleven recurring solar
 jets originated from two different sites  (site 1 and site 2)
close to each other ($\approx$ 11 Mm) in the NOAA active region (AR) 12035 during 15--16 April 2014.
 The jets were observed by the {\em Atmospheric Imaging Assembly} (AIA) 
telescope onboard  the {\em Solar Dynamics Observatory} (SDO) satellite.
 Two jets were observed by the {\em Aryabhatta Research Institute of Observational Sciences} (ARIES), Nainital, India telescope in  
H$\alpha$. 
On 15 April flux emergence is   important   in site 1 while on 16 April flux emergence and  cancellation mechanisms 
are involved in both sites.
The jets of both sites have parallel trajectories and move to the
 south with a speed between 100 and 360 km s$^{-1}$.
 The jets of site 2 occurred during the second day and have a tendency to move towards the jets of site 1 and merge with them.
We conjecture that the slippage of the jets could  be explained by the 
complex topology of the region with  the presence of a few low-altitude 
null points 
and many quasi-separatrix layers ({\em QSLs}), which  could intersect
with one another.
\end{abstract}

\end{opening}

\section{Introduction}
     \label{S-Intro}

Solar jets are small scale dynamic events observed as collimated
ejections of plasma material
from the lower solar atmosphere to coronal heights. 
Many jets are observed in active regions 
 (for example see \opencite{Schmieder1983},  \opencite{Gu1994}, 
\opencite{Schmieder13},
\opencite{Guo13}, \opencite{Chandra17} and references cited therein).  
They are observed in a wide range of electromagnetic spectra from the
 optical to UV--EUV and  X-rays. 
Jets are particularly  visible  in coronal hole (CH) 
regions \citep{Madjarska}, 
probably because these are regions of open magnetic fields  with less EUV background emission. 
In chromospheric  spectral lines, solar 
jets are called surges. Surges and jets are associated with base brightening, referred as ``jet bright points''
 \citep{Sterling16}. For state-of-art observations, theory and 
models of jets, we refer the review of \inlinecite{Raouafi16}.

The soft X-ray telescope onboard {\em Yohkoh} satellite opened a new view on  understanding solar 
jets with the pioneer paper of \inlinecite{Shibata92} followed by  several studies 
deriving the characteristics 
of solar jets such as 
velocity, height, temperature, width, and density \citep{Shimojo96,Shimojo98,Shimojo00,Schmieder13,Chandra15}. 
Using the SECCHI/EUVI data \inlinecite{Nistico09}
 performed a morphological study of 79 coronal jets and found that 
 31 have helical structures. Five jets were associated with micro--CMEs. 

In some cases  solar jets are associated with 
 eruption  of the full 
  base region.
Such types of jets are called 
``blow-out jets'' \citep{Moore10,Moore13,Hong11,Shen12,Chandra17}. When  successive jets occur at the same 
location and with similar morphological properties in some time interval, they are called 
homologous recurrent jets. Usually they are ejected in the same direction and their footpoint structures are similar.
Several studies have been performed about recurrent homologous jets in different wavelengths such as 
in H$\alpha$ \citep{Asai01,Uddin12,Chandra17},  in EUV \citep{Chae99,Jiang07,Schmieder13,Joshi17}, and in X-rays \citep{Kim01,Kamio07,Sterling16}. 

Based on the X-ray jets  and EUV/UV jets observed by {\em Hinode} and 
 SDO/AIA respectively, a new jet formation model has been proposed by
Sterling and co-workers \citep{Sterling17,Panesar17}. According to their model all the jets would be 
originated by the eruption of a small--scale filament or ``minifilament", as already mentioned in
\inlinecite{Hong11} and \inlinecite{Shen12}.
 \inlinecite{Sterling16} found that the minifilament eruption kinematics
was similar to the kinematics of most observed large filament eruptions. In large filament
 eruptions it  was commonly observed that the slow rise phase is followed by the fast rise
acceleration phase. The association of minifilament eruptions was also observed
in CH regions \citep{Sterling15} and in quiet regions \citep{Panesar17}.
However, \inlinecite{Sterling16} and \inlinecite{Sterling17} found that the existence of
minifilament eruptions in active region jets are sometimes difficult to observe.

The explosive dynamic nature, morphology and 
magnetic configuration of the jets at their base 
support the idea that they are the result of magnetic reconnection. 
Magnetic reconnection at null--points is proposed in several jet models. 
It is also confirmed by magnetic field extrapolations  and simulations
\citep{Moreno08,Zhang12,Schmieder13}. In some of the studies circular ribbons have been 
observed at the base of  jets \citep{Wang12}. This topology also 
supports the presence of magnetic null--points above  jet locations.

 However, using  photospheric magnetic field extrapolations, 
\inlinecite{Mandrini1996} and  \inlinecite{Guo2013} explained the 
jets by the presence of bald patch (BP) regions along 
 QSLs  without  the existence of null--points.  QSLs 
are thin layers with a finite gradient in the
 connectivity of the magnetic field where magnetic reconnection can occur.
Very recently, \inlinecite{Chandra17} computed  the magnetic topology of  
the NOAA active region 10484 on 21--24 October 2003 and
 found that the flares and jets  occurring in this region were due to
magnetic reconnection at the BP separatrices.
 BP separatrices  are regions 
where magnetic field lines  are not anchored on the
 photosphere but are tangent  between two  
different magnetic regions. 
The global magnetic topology  of active regions  is commonly determined by  
potential extrapolation which allows the computation of QSL locations, which are 
relatively robustly determined \citep{Demoulin1996}.
QSLs are measured by the computation of the squashing degree {\em Q}, proposed by \inlinecite {Titov2002}.
The squashing degree increases with the growth of connectivity gradients, and
becomes infinite for separatrices.
QSLs footprints in the photosphere coincide with the flare
 ribbons even if the shape is not always perfectly represented \citep{Dalmasse2015,Zuccarello17}.

Flux emergence in the photosphere could  be one of the drivers of solar jets. 
According to this scenario, it is the reconnection between the 
newly emerged magnetic flux (EMF) and the pre--existing magnetic flux that leads to the formation of jets.
 The reconnection can 
be driven by the motions in the newly emerged flux. To explain this magnetic 
reconnection the first dynamical model was proposed in two dimensions (2D) 
\citep{Yokoyama95,Yokoyama96}. 
Now EUV and X-ray jets are being modeled with 3D magnetohydrodynamic (MHD) simulations 
\citep{Moreno08,Moreno13,Archontis13,Torok09}. The background corona 
  for both 2D and 3D models is an open magnetic field.

Another scenario proposed for the jet origin is  the 
loss--of--equilibrium mechanism. In this mechanism, the jet 
occurs when the stressed closed flux under the 
null--point reconnects with the surrounding quasi-potential flux exterior to  the fan surface.
 The reconnection starts when some 
threshold altitude is reached. In this scenario, the jet displays an untwisting 
motion \citep{Pariat09,Pariat10,Dalmasse12,Pariat15}.   

Flux cancellation at the jet locations is also 
frequently proposed as the driver of jets
\citep{Innes10,Liu11,Innes16,Adams14,Young14,Cheung15,Chen15}. \inlinecite{Zhang00} proposed such flux cancellation
 between oppositely directed magnetic 
field  to explain macrospicules and microscopic jets.
\inlinecite{Adams14} also found  magnetic flux convergence and cancellation along the polarity
inversion line (PIL), where the jets were initiated. 
In \inlinecite{Guo13} the cancelling flux occurred at the edge of EMF 
during its expansion. Therefore, it is still 
discussed if jets are due to flux emerging or cancelling or both.  

The NOAA AR 12035 had a very high productivity of activity
 during 15--16 April 2014
with  many confined and eruptive 
flares in  its northern part and jets in its southern part. The confined and eruptive flares from this active region have been  studied in 
\inlinecite{Zuccarello17}. They explained that the eruptive flares become more and more confined, when the overlying magnetic field becomes
less and less anti--parallel.   

In this study, we investigate the morphology,
 kinematics and magnetic causes of the solar jets 
during 15--16 April 2014 from NOAA AR 12035.
 The paper is organized as follows.
Section \ref{obs} presents the observational data, morphology, 
kinematics and the magnetic configuration of the active region. 
The magnetic topology of the active region is discussed in 
Section \ref{mag_top}. Finally, in Section \ref{result} we discuss and conclude our results.

\section{Observations}
     \label{obs}
\begin{figure}[t]
\vspace*{-0.4cm}
\hspace{-1.2cm}
\includegraphics[width=1.08\textwidth, clip=]{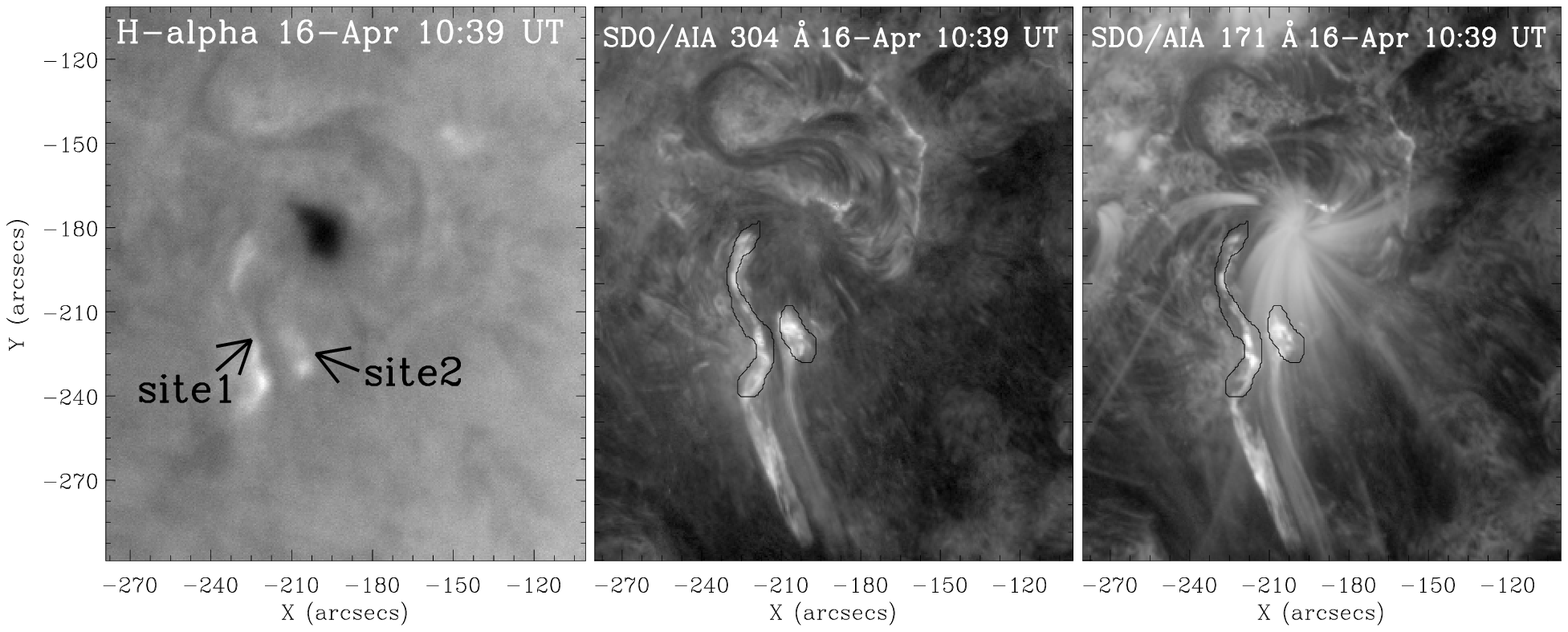}
\vspace*{-0.4cm}
\caption{Jets J${_5}$  and J\ensuremath{'}${_5}$  from site 1 and site 2 
respectively  in the  south  of the main sunspot  of AR 12035 on 16 April 2014 {\it i.e.} from left to right  in H$\alpha$, observed in Nainital,
India (ARIES),
 in 304 \AA\ and in 171 \AA\ observed  with SDO/AIA.
The contours of H$\alpha$ are over-plotted on to 304 \AA\ and 171 \AA.
 The jets look to be overlapping according to the scale resolution of the images. }
\label{fig:hal}
\end{figure}

The NOAA AR 12035 produced many solar jets during 
15--16 April 2014 towards the south direction. We have selected
eleven clearly visible jets for our investigation. 
Their description is given in Table 1. 
These jets were observed by the  {\em Atmospheric Imaging
Assembly} (AIA, \opencite{Lemen12}) onboard  {\em Solar Dynamics Observatory} (SDO,  
\opencite{Pesnell12}) in different 
EUV and UV wavebands. 
 The spatial and temporal resolution of AIA 
data are 1.2$^{\arcsec}$ and 12 s respectively. 
To study the magnetic configuration around the location of the 
jets we have used the line--of--sight (LOS) magnetic field data from the 
{\em Heliospheric Magnetic Imager} (HMI, 
\opencite{Schou11}) having a spatial resolution of 1$^{\arcsec}$ 
and  a cadence of 45 s.
On 16 April 2014  two jets, that we will name jet J${_5}$ 
and J\ensuremath{'}${_5}$ in the next section, 
 were observed in H$\alpha$ by the 15 cm Coud\'e telescope
operating at Aryabhatta Research Institute of Observational Sciences (ARIES), Nainital, India (Figure~\ref{fig:hal}). 
The pixel size and cadence are  0.58$^{\arcsec}$ and 1 min respectively. 
For consistency, we have shifted all the images to 16 April 2014 10:30 UT.

For identifying the onset and peak time of the jets,
we look into the temporal evolution of intensity at the jet foot-point.
We create a box containing the bright jet base and calculate 
the total intensity inside it. Then this total intensity is 
normalized by the intensity of the quiet region.

The morphology, kinematics and magnetic configuration of the jet locations
are described in the following subsections.

\subsection{EUV Observations}
\label{obs1}

\begin{figure}
\includegraphics[width=1.0\textwidth, clip=]{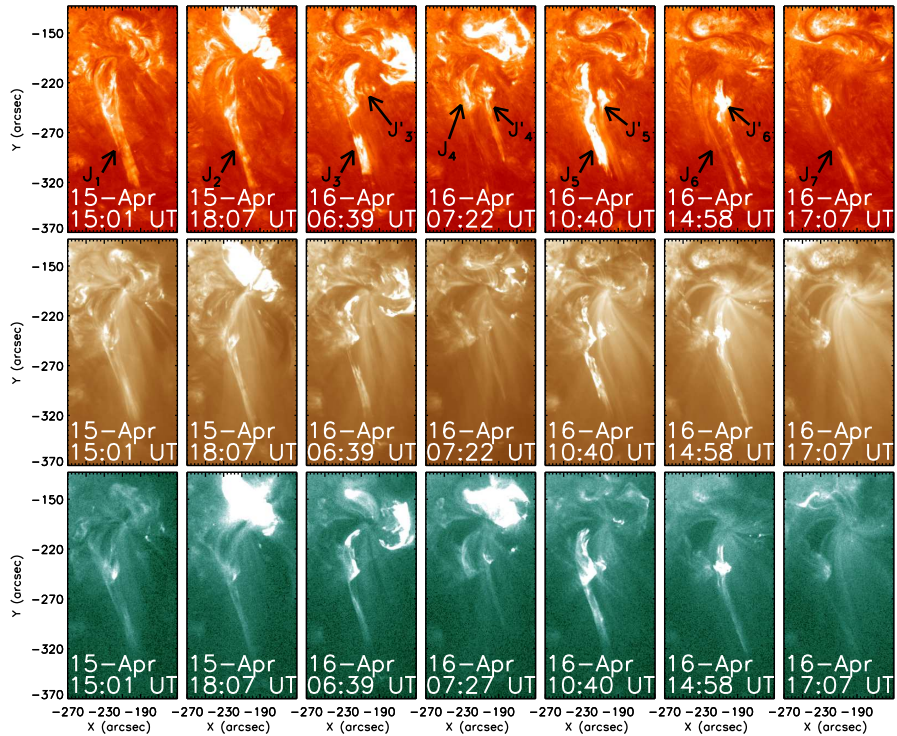}
\caption{Recurrent jets in the active region AR 12035  on 15 and 16 April 2014.
Upper, middle and bottom panels are images of AIA
filters at   304 \AA, 193 \AA,  and 94 \AA\ respectively.  
The arrows indicate the bright emission of the  jets in two different locations mentioned as site 1 and site 2 in Figure~\ref{fig:hal}.}
\label{fig:jet/morpho}
\end{figure}

The eleven selected jets during 15--16 April 2014
are named as J${_1}$--J${_7}$ and  J\ensuremath{'}${_3}$-- J\ensuremath{'}${_6}$ 
 respectively.
Out of eleven jets, the first two jets J${_1}$ and J${_2}$ occur on 15 April and the remaining nine jets 
are on 16 April 2014.
These jets originated from two locations in the south part of the NOAA AR 12035.
One location is at position [X,Y] = [-220$^{''}$, -215$^{''}$] and the other is at 
 [-205$^{''}$, -215$^{''}$] (Figure~\ref{fig:hal} left panel). 
In this paper, we will refer them {\bf as} 
site 1 jets (J) and  site 2 jets (J\ensuremath{'}) respectively.  The two sites are at a distance 
of 15 arcsec from each other ($\approx$  11 Mm).
Figure~\ref{fig:jet/morpho} displays images of the eleven jets observed  
with the AIA filters at 304 {\AA} (top), 193 {\AA} (middle) 
and 94 {\AA} (bottom) during their peak phase.
 Almost all the jets are visible in all  EUV channels, which 
indicates the multi-thermal nature of the  jets. 
The onset and end time of each jet are summarized in Table \ref{recurrent}.

\begin{figure}[t]
\includegraphics[width=1.0\textwidth, clip=]{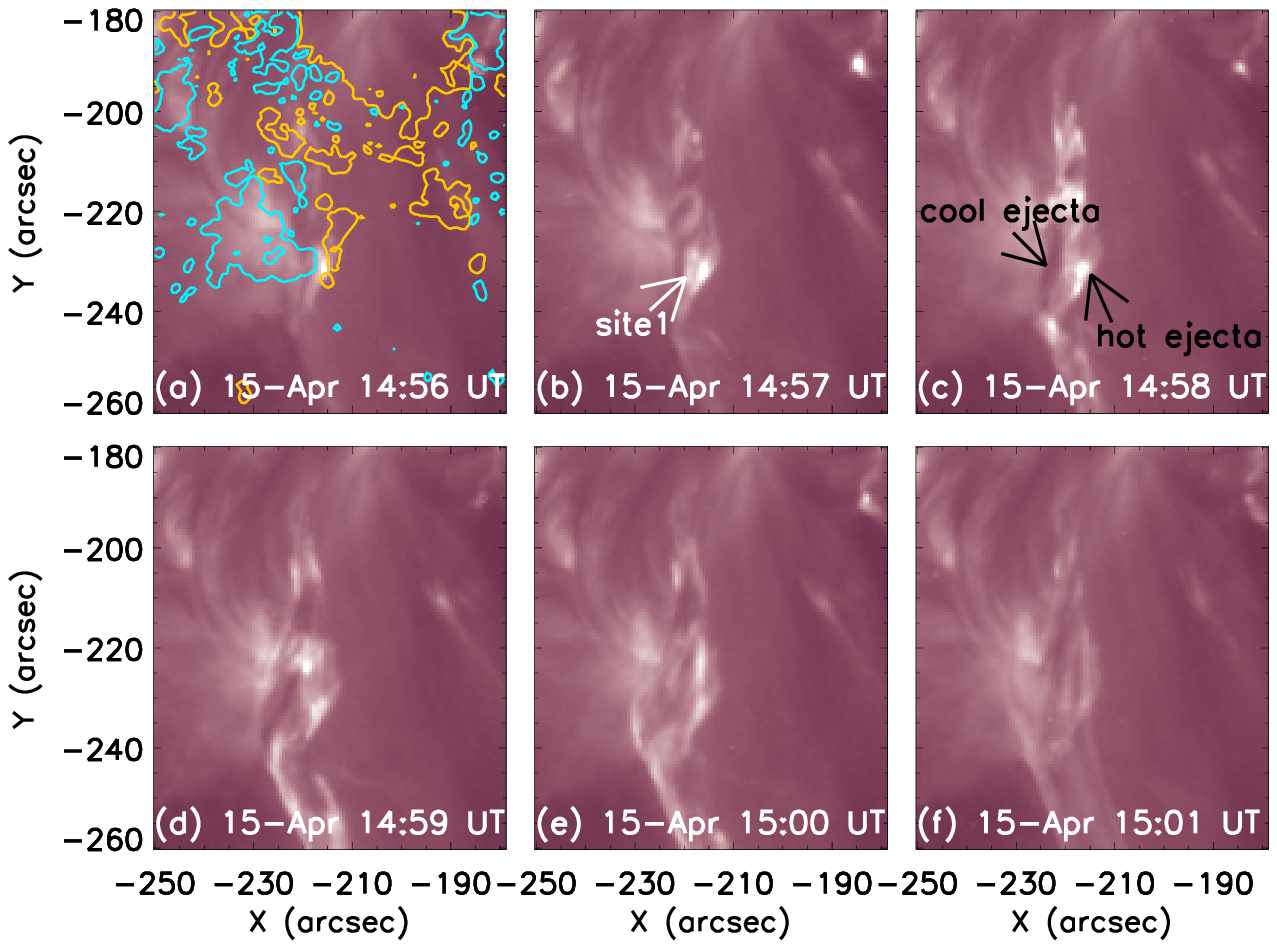}
\caption{Evolution of jet J${_1}$  initiated in site 1 
observed by AIA 211 \AA. The contours on the first image are HMI 
longitudinal magnetic field with
 yellow and cyan colors for positive and negative polarities respectively.
The contour levels are $\pm$100 Gauss. 
The site 1 is indicated by  a white arrow. 
The cool and hot part of the ejected jet J${_1}$ are indicated 
 by black  arrows.}
\label{fig:jet1}
\end{figure}

Jet J${_1}$ started on 15 April 2014 with a bright base at site 1. 
The zoomed view of evolution of jet 
J${_1}$ in AIA 211 \AA\ is shown in Figure~\ref{fig:jet1}. 
Together with the ejection of bright material, we
 have also observed the ejection of cool and 
dark material in 211 {\AA} (see Figure~\ref{fig:jet1}c). 
 The dark jet is due to the presence of cool material
   absorbing the coronal emission. The bright
and cool jet material were rotating clockwise (see the attached movie in 211 \AA).
After the onset of $\approx$ 2 min it started to rotate anti-clockwise. 
This indicates the untwisting of the system to relax to a lower energy state by propagating 
twist outwards. 
The jet J${_2}$ started also with a bright base similar to
 J${_1}$ from the same location.
This jet was also showing untwisting like as in the  case of 
J${_1}$. However, its base was less bright and less 
broader than J${_1}$. 

\begin{figure}[t]
\includegraphics[width=1.0\textwidth, clip=]{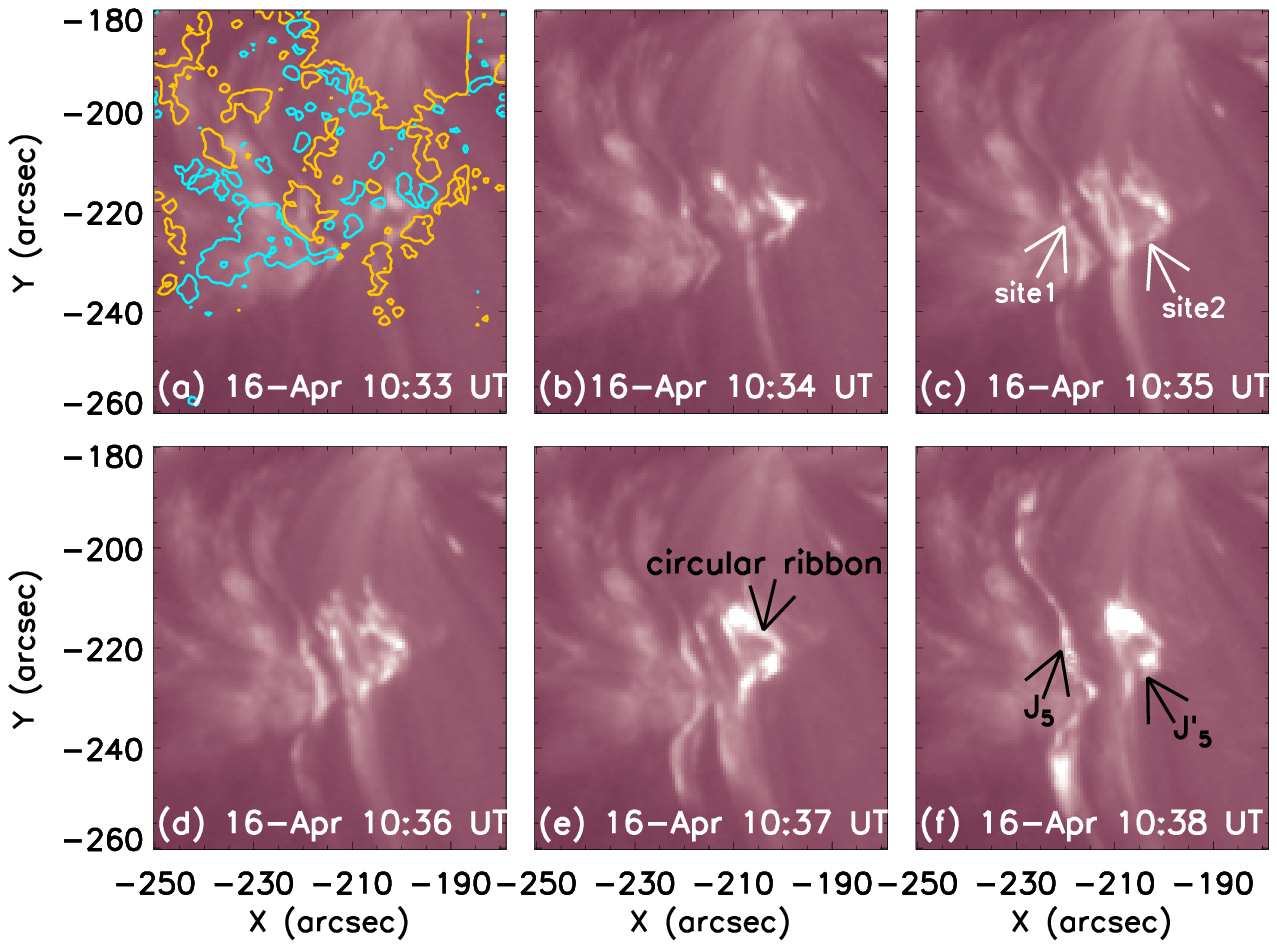}
\caption{Evolution of jets J${_5}$ and J\ensuremath{'}${_5}$  
from site 1 and site 2 (indicated by white arrows) 
observed by AIA 211 \AA. The contours on the first image are HMI 
longitudinal magnetic field.
 Yellow and cyan colors are for positive and negative polarities respectively. 
The contour levels are $\pm$100 Gauss. The positions of jets J${_5}$, J\ensuremath{'}${_5}$ and  the
circular ribbon are indicated  by black arrows.}
\label{fig:jet5}
\end{figure}

\begin{figure}[t]
\vspace*{-0.1cm}
\hspace{-0.5cm}
\includegraphics[width=1.0\textwidth, angle=0, clip=]{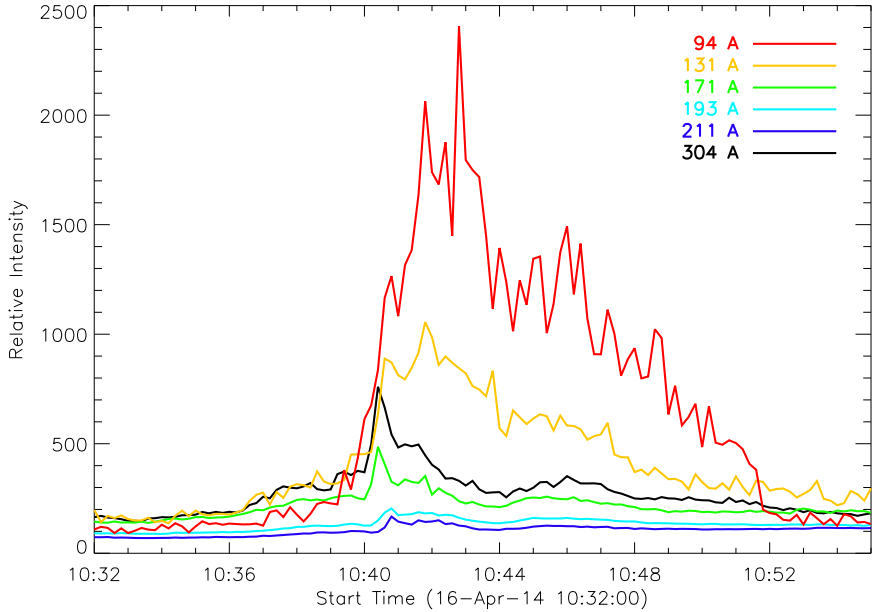}
\vspace{-0.2cm}
\caption{Intensity profile of jet J${_5}$ at base location. Different colors
represent different SDO/AIA wavelengths. The peak of the cooler component is 
earlier (304 \AA: longer wavelength) than the peak of the hotter component (94 \AA: shorter wavelength).}
\label{fig:inten}
\end{figure}

\begin{figure}[t]
\includegraphics[width=0.70\textwidth, clip=]{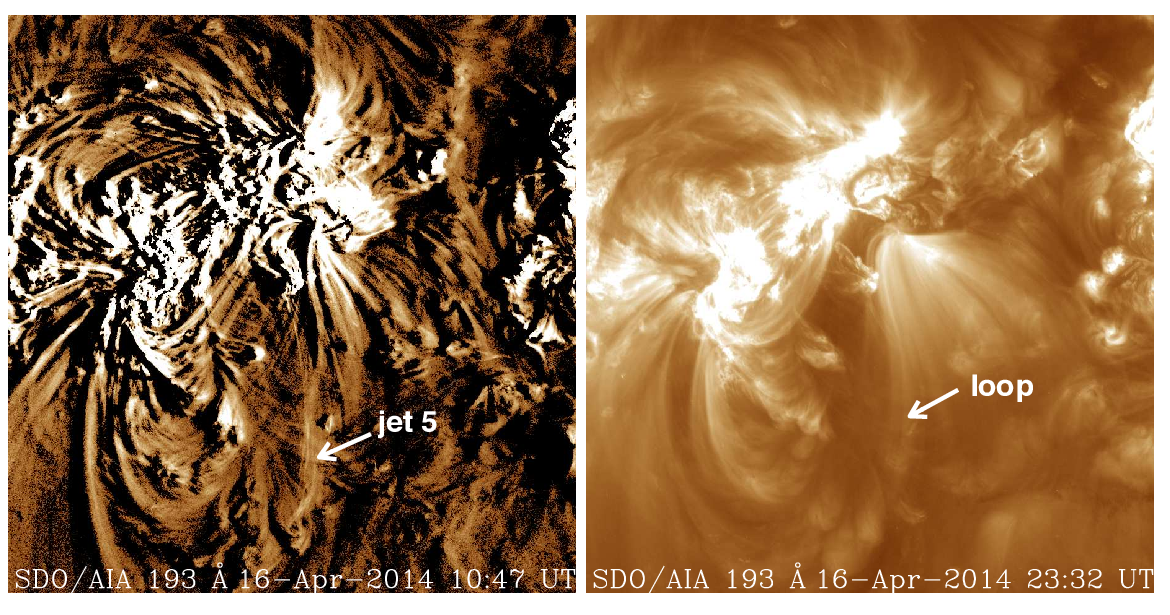}
\caption{Sigmoidal orientation of the  loops in  the active region 
in the AIA  193 \AA\, filter (right panel) and in the difference image (left panel).
The jet J$_5$ follows the path of the loop.} 
 \label{fig:loop}
\end{figure}

On 16 April 2014 J${_3}$ and J\ensuremath{'}${_3}$
 started from  site 1 and site 2 respectively. 
The base of J\ensuremath{'}${_3}$ was like a circular ribbon and broader than J${_3}$.
Jet J\ensuremath{'}${_4}$  is bigger than J${_4}$. 
Jet J${_4}$ peaks almost simultaneously in all EUV wavelengths around 07:17 UT. 
The peak time for J\ensuremath{'}${_4}$ is five minutes later than  J${_4}$, around 
07:21 UT in all wavebands. One difference between 
these jets and  J${_1}$--J\ensuremath{'}${_3}$ was that they have no rotation.
The evolution of jets J${_5}$ and J\ensuremath{'}${_5}$ is presented in 
Figure~\ref{fig:jet5} in AIA 211 \AA. 
In contrast to the other jets, the peak time for jet J${_5}$ and J\ensuremath{'}${_5}$
 is different for different wavebands. The peak
 time for J${_5}$ at 304 \AA\ is 10:38:30 UT, 
and at 94 \AA\ is 10:41 UT. For the case of J\ensuremath{'}${_5}$,
the peak time in 304 \AA\ is 10:37:30 and at 94 \AA\ is 10:38 UT. 
For these jets, the peak for
cool plasma (longer wavelength 304 \AA) appears earlier than the hot plasma 
material (shorter wavelength 94 \AA).
The temporal evolution of flux at the base of jet J$_5$ is shown in  
Figure~\ref{fig:inten}.
This behavior is contrary to  other jets reported in the literature
 where hotter plasma appears before cooler plasma,
 suggesting some mechanism of cooling versus time \citep{Alex99}.
These are the largest jets among the ones discussed in this study. As the event
progress, interestingly part of jet J\ensuremath{'}${_5}$  was detached from it and 
moved towards  site 1 and finally merged with jet
J${_5}$ (see the attached movie in 211 \AA). 
We have estimated the speed of
 J\ensuremath{'}${_5}$ towards J${_5}$ as $\approx$ 45 km s$^{-1}$.
Merging of the broken part of J\ensuremath{'}${_5}$ 
with J${_5}$ made it bigger as it was ejected in the south 
as well as in the north direction at the same time. 
We have also noticed the circular ribbon at the base of 
J\ensuremath{'}${_5}$, shown in Figure~\ref{fig:jet5}e.
Looking at the evolution of the J\ensuremath{'}${_5}$ jet, we found
that the jet follows a sigmoidal-shape loop path visible in AIA 193 \AA\ 
(see Figure~\ref{fig:loop}).
 These loops are originating from the sunspot and going towards  the south  with a 
sigmoid-shape. The sigmoidal-shape of these loops could be due the the clockwise rotation of big positive 
polarity spot (see Section 2.4). 
We have observed the clockwise rotation in jet  
J${_5}$ and the calculated rotation speed in four wavelengths
171 \AA\ , 193\AA\ , 211 \AA\ and in 304 \AA. 
The speed varies from 90 km s$^{-1}$ to 130 km s$^{-1}$.
Jet J${_6}$ was small and of weak intensity whereas J\ensuremath{'}${_6}$ had
 a strong intensity and was very bright, wide and had a circular base. 
J\ensuremath{'}${_6}$ started to move towards site 1, but like in the case of 
J\ensuremath{'}${_5}$, it could not reach up to J${_6}$ location. We did not find any rotation in this jet. 
Jet J${_7}$ started from site 1 location and  also showed clockwise rotation.

Let us summarize the different morphological features of these jets.
 We have found that all jets from  site 1 have similar morphology.
Jets  from site 2 also have similar morphology, but this
is different from the morphology of jets ejected from site 1.
One common feature in all jets of site 2 on 16 April
was that after their trigger they all have a tendency to 
move towards site 1 before or during the ejections. 
 It seems that for the case of J${_5}$ and  J\ensuremath{'}${_5}$, 
there is a connection between these two.
 Another common feature of jets 
originated from site 2 is 
that they all follow the sigmoidal  
loops visible in different AIA wavebands.  
The jets from site 2 occur before or almost simultaneously 
(in the case of J\ensuremath{'}${_4}$) with the jets from site 1.
This is true apart from the jet J$_6$. The main difference is 
that J$_6$ is quite weak among all of the coupled jets.
We have noticed that jets observed in 193 \AA\ 
are thinner than in 304 \AA. All jets started
with a bright base, similar to  common X-ray jet observations.

\subsection{H$\alpha$ Observations}
     \label{obs2}

Two jets-- jet J${_5}$ and J\ensuremath{'}${_5}$ were also observed 
in the H$\alpha$ line center (6563 \AA) by ARIES, 
Nainital, India as a bright ejecta. The bandwidth of the H$\alpha$ 
filter was 0.5 \AA\ with a cadence of 1 min.
H$\alpha$ jets from both jets sites started at  $\approx$ 10:35 UT,
 one minute later than  in the  EUV wavebands and they faded
away  at $\approx$ 10:50 UT. We also observed the dark 
material ejection surge between the bright jets from the two 
sites site 1 and site 2 respectively. 
To compare the spatial location of H$\alpha$ with EUV images,
 we have over-plotted the
contours of H$\alpha$ in AIA 304 \AA\ and 171 \AA\ . 
The result is shown in Figure~\ref{fig:hal}. 
We found  that H$\alpha$ jets are coaligned with the EUV jets.
 However, the position of the H$\alpha$ jet is 
only at the origin of the EUV jets. It may be because the H$\alpha$ jets 
are in  the chromosphere and have 
less height than the EUV jets. The bright
 ejections in H$\alpha$ are followed by dark material ejections. 
It could be the cooler part of the jet. 
This cooler part is spatially shifted towards the west side from the bright jets.

\subsection{Height--Time Analysis}
     \label{obs3}

\begin{figure}[t]
\hspace{-0.8cm}
\includegraphics[width=1.06\textwidth, clip=]{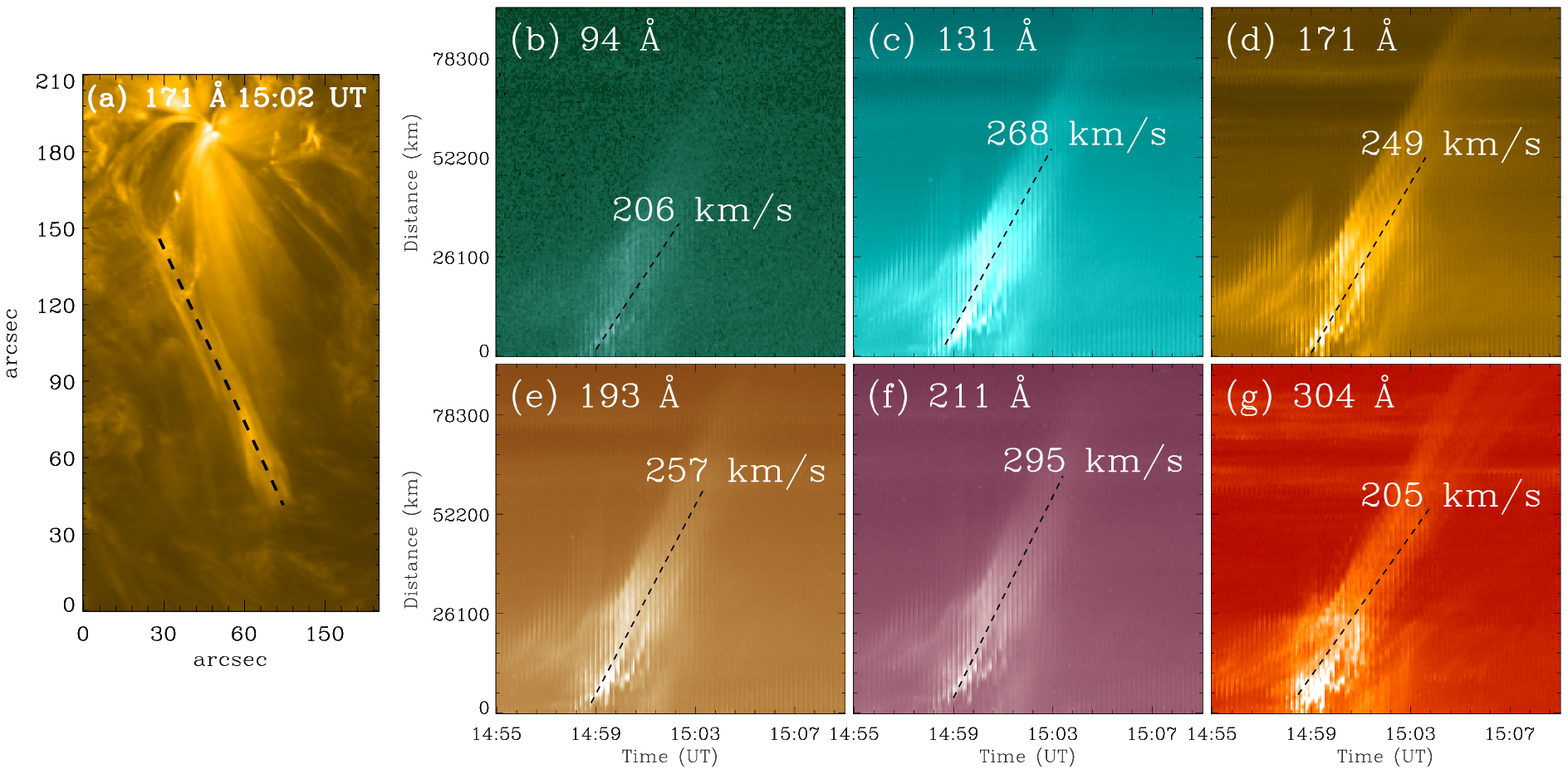}
\vspace{-0.6cm}
\caption{Time-slice analysis of jet J${_1}$ on 15 April 2014. Left: position of slit along the jet, 
right: velocity comparison at different wavelengths.}
\label{fig:slice1}
\end{figure}

\begin{figure}[h]
\hspace{-0.8cm}
\includegraphics[width=1.00\textwidth,clip=]{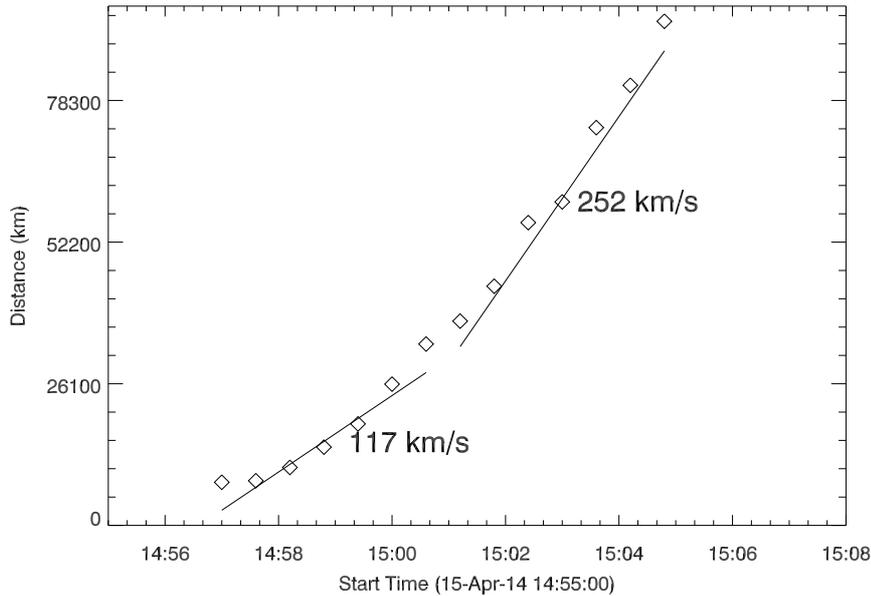}
\caption{An example of a   height--time plot for Jet J${_1}$ derived using the
leading edge procedure.}
\label{fig:leading_edge}
\end{figure}

To understand the kinematics and dynamics of the observed jets, we have made a time-distance analysis in 
different AIA/EUV channels {\it i.e.} 94, 131, 171, 193, 211, and 304 \AA. 
To perform this analysis, we adopted two methods, one is the time--slice technique and 
the other is tracking of the leading edge of the jets. For the time--slice technique, we 
fixed a slit at the center of the jets and observed the motion of plasma along the slit. 
An example of this time--slice analysis of jet J${_1}$ on 15 April 2014  is presented in
 Figure~\ref{fig:slice1}.
In the figure, the left image shows the location of  the slit (drawn as a black dashed line) and the right 
denotes the height-distance maps at different EUV wavelengths. 
Using this time--slice technique, we have computed the heights and projected
 speeds
 for the different structures of all jets at different wavelengths,
shown in the right panel of the figures.
For the same jet of 15 April 2014 at 193 \AA, the height--time plot using the leading 
edge procedure is presented in  Figure~\ref{fig:leading_edge}. 
The curve  has an exponential behavior with  two  acceleration phases.
 We decided to fit  linearly  the beginning of the expansion and then 
the later phase in order to compared the values with those of the other 
methods. Two speed values are derived. The first is 117 km s$^{-1}$ which is 
nearly half of the speed derived  after the acceleration phase 
(252  km s$^{-1}$). The  error is around 10  km s$^{-1}$ according to the points chosen 
for the fits.


The time--slice technique indicates that every jet has
 multi-speed structures 
and the speeds of the jets are different in different EUV wavelengths.
The average speed by the time--slice technique varies in different wavelengths
 for J${_1}$: 205 -- 295 km s$^{-1}$, J${_2}$: 
177--235 km s$^{-1}$, J${_3}$: 132--163 km s$^{-1}$, J\ensuremath{'}${_3}$: 100 --121 km s$^{-1}$, 
J${_4}$: 133 --164 km s$^{-1}$, J\ensuremath{'}${_4}$: 295 --325 km s$^{-1}$, J${_5}$: 174 --202 km s$^{-1}$,
J\ensuremath{'}${_5}$: 275 --364 km s$^{-1}$, J${_6}$: 156 --197 km s$^{-1}$  
J\ensuremath{'}${_6}$: 304 --343 km s$^{-1}$, and for J${_7}$: 154 --217 km s$^{-1}$ respectively.
The dispersion of the values for one jet according to the  different considered wavelengths is 
between 10$\%$ to 50$\%$, with no rule.
 It is difficult to understand
if it is just the range of the uncertainties of the measurements   or  if it really
 corresponds to the existence of multi-components in the jet  with multi-temperatures and speeds 
or if the slit of the time distance diagram is crossing different components of the jets.
We have compared these results with the speed derived by the leading edge procedure.
 We found  that the fast speed fits with the time-distance derived 
velocity. The time-distance technique with a straight slit ignores the 
first phase of the jets.  The values of speed derived by both methods 
are presented in Table \ref{recurrent}.

We have also computed the lifetimes, widths, and the maximum heights 
of each jet.
The lifetimes vary from 6 to 24 min.
 J${_5}$ has the maximum lifetime (24 min), whereas J\ensuremath{'}${_4}$ 
has the minimum (6 min) lifetime.
In general, we found the narrow and long life--time jets achieved 
more height than the wide and short lifetime jets.
We have also noticed that the lifetimes of jets are longer in 304 \AA\ and in 171 \AA\ than in 94 \AA. 
The width ranges from 2.6--6 Mm. 
The maximum height was attained by J${_5}$ and it was 217 Mm.

\subsection{Evolution of Photospheric Magnetic Field}
     \label{obs3}

\begin{figure}[t]
\includegraphics[width=1.00\textwidth,angle=0,clip=]{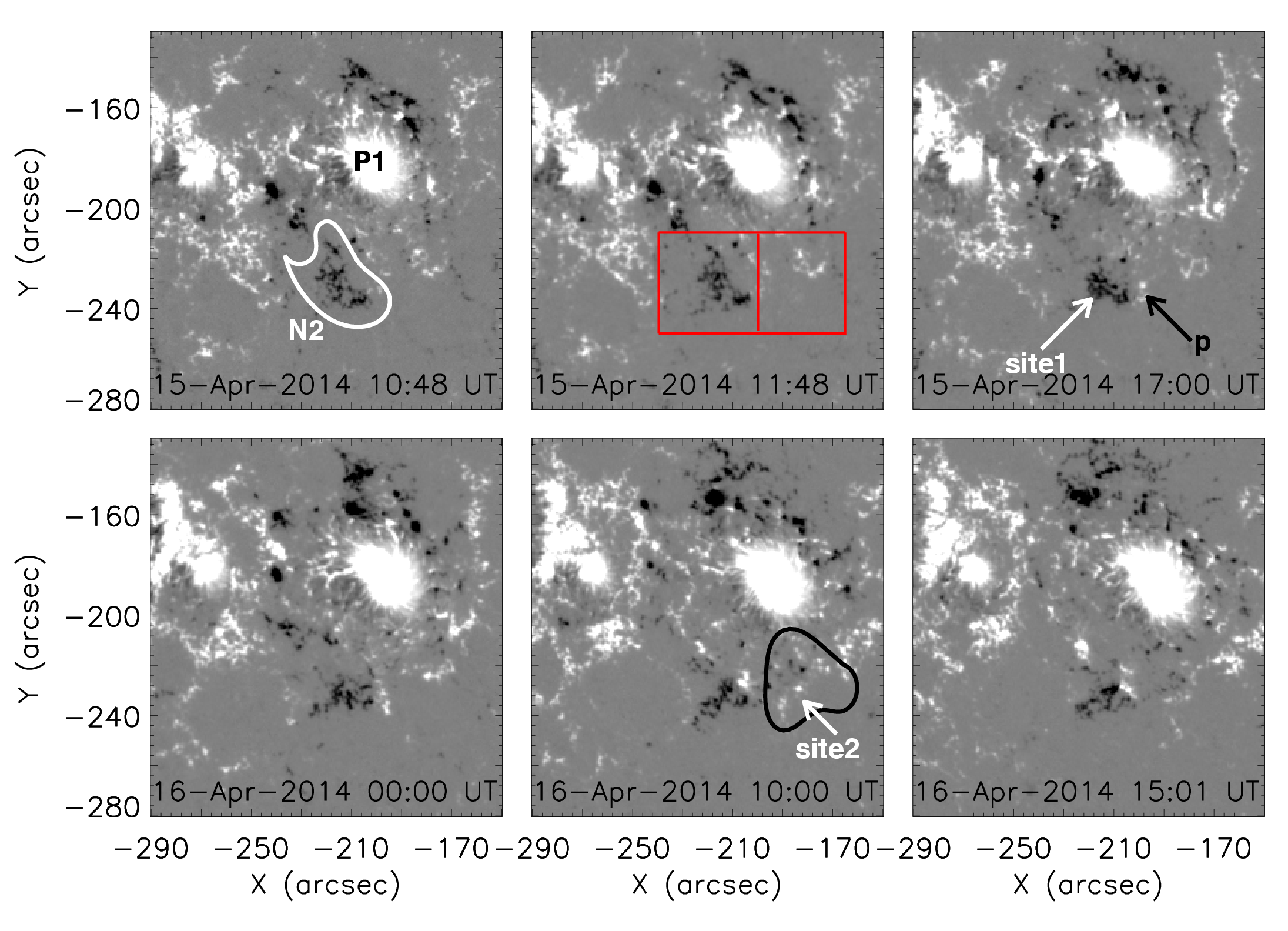}
\caption{ Magnetic field evolution of NOAA active region 12035. The two jet
locations  (site 1 between N2 and p, site 2 at the edge of the supergranule 
indicated by the black contour) are  indicated by white arrows.
The red boxes in the left of the bottom panel show the site 1 (left) and 
site 2 (right) locations respectively and the field of view of Figures \ref{fig:mag1}
and \ref{fig:mag2}.} 
\label{fig:mag}
\end{figure}

The NOAA AR 12035 appeared at the east limb on
 11 April 2014  with  a $\beta$ magnetic 
configuration. It went over the west limb on  
24 April 2014. Figure~\ref{fig:mag} presents the magnetic field evolution of 
the active region observed by SDO/HMI during 15--16 April 2014.
The active region  consists of two large positive polarity sunspots P1 and P2 followed by the negative 
polarity spot N1 (see Figure~\ref{fig:top}a).
The positive polarity P1 behaves like a decaying sunspot with a ``moat region"
 around it. When the 
sunspot is close to its decay stage, it loses its polarity by dispersing
 in all directions. This dispersed positive polarity
 cancels a part of the negative magnetic polarity 
left at the site of the jet location.
The origin of jet activity lies between the
  two leading positive polarities. As seen in 
Figure~\ref{fig:mag} the whole active region shows 
clockwise rotation (see also the attached movie MOV3).
Together with the rotation of the whole active region, the 
western big spot P1 also shows a rotation
 in the same direction as the active region.

The sunspot rotation makes the active region
 sheared and the loops connecting the preceding
 positive polarity and the
following negative polarity become sigmoidal 
 (see Figure~\ref{fig:loop}). The jets are ejected in the 
south-east direction instead of south direction probably because of the 
upper part of the sigmoidal loops.

To investigate the magnetic field evolution at the jet's origin, we have 
carefully analyzed the development of different polarities. 
On 15 April, we observed the positive polarity P1 and in its south a bunch of 
negative polarity N2 and  a positive polarity p (Figure~\ref{fig:mag}).  
Site 1 of the jet's origin is located between N2 and p. 
Site 2 is located to the west of site 1.

\begin{figure}[t]
\includegraphics[width=0.70\textwidth,angle=90,clip=]{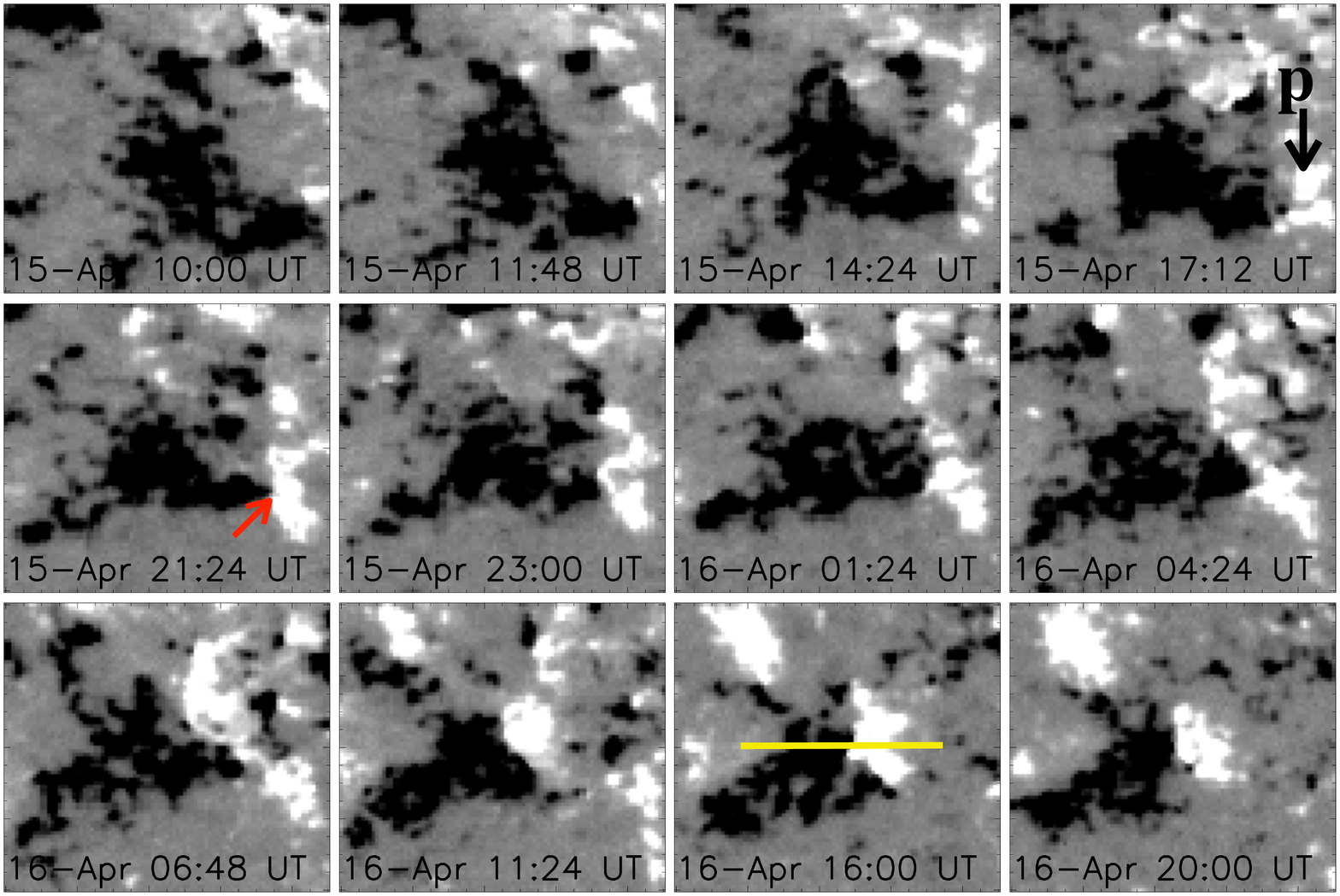}
\caption{Evolution of magnetic field at the site 1 location. 
The flux cancellation at site 1 between the negative polarity and the positive
polarity (p) is shown by a red arrow.
The horizontal yellow line indicates the location of the slit, where the time-distance evolution of flux is 
calculated (see Figure \ref{fig:mag_slice}).} 
\label{fig:mag1}
\end{figure}

\begin{figure}[t]
\includegraphics[width=1.0\textwidth,angle=0,clip=]{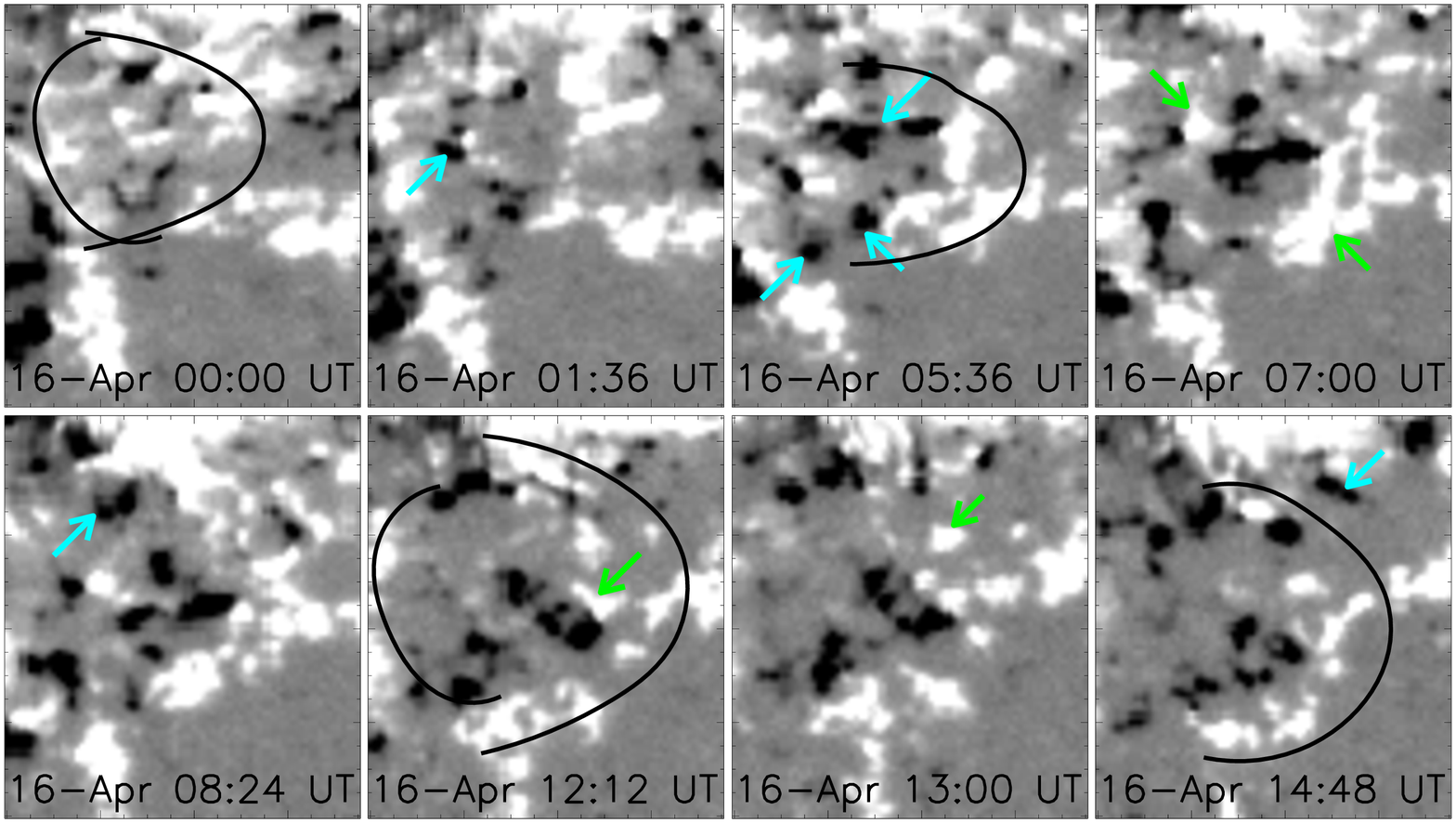}
\caption{Evolution of magnetic field at the site 2 location. The positive (negative) flux emergence is shown by green (cyan)
arrows respectively. The roundish black curve indicate kind of supergranule cell.}
\label{fig:mag2}
\end{figure}

\begin{figure}[t]
\includegraphics[width=1.00\textwidth,clip=]{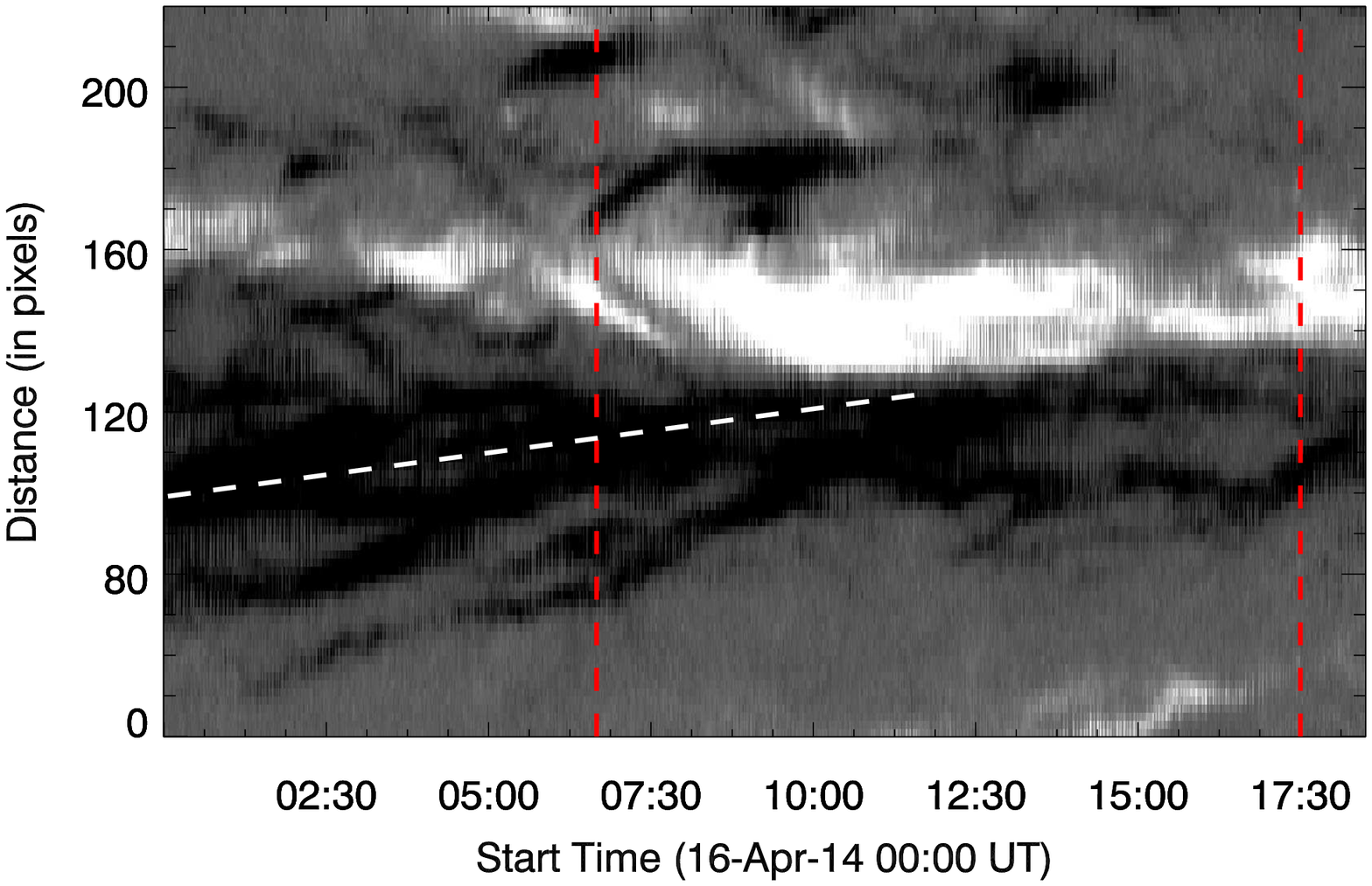}
\vspace*{-4.5cm}
\caption{Magnetic field evolution along the slice shown in Figure 9 at site 1. Two vertical lines in the image 
indicate the time of jet activity.}
\label{fig:mag_slice}
\end{figure}

\begin{figure}[t]
\vspace*{0.2cm}
\includegraphics[width=1.00\textwidth,clip=]{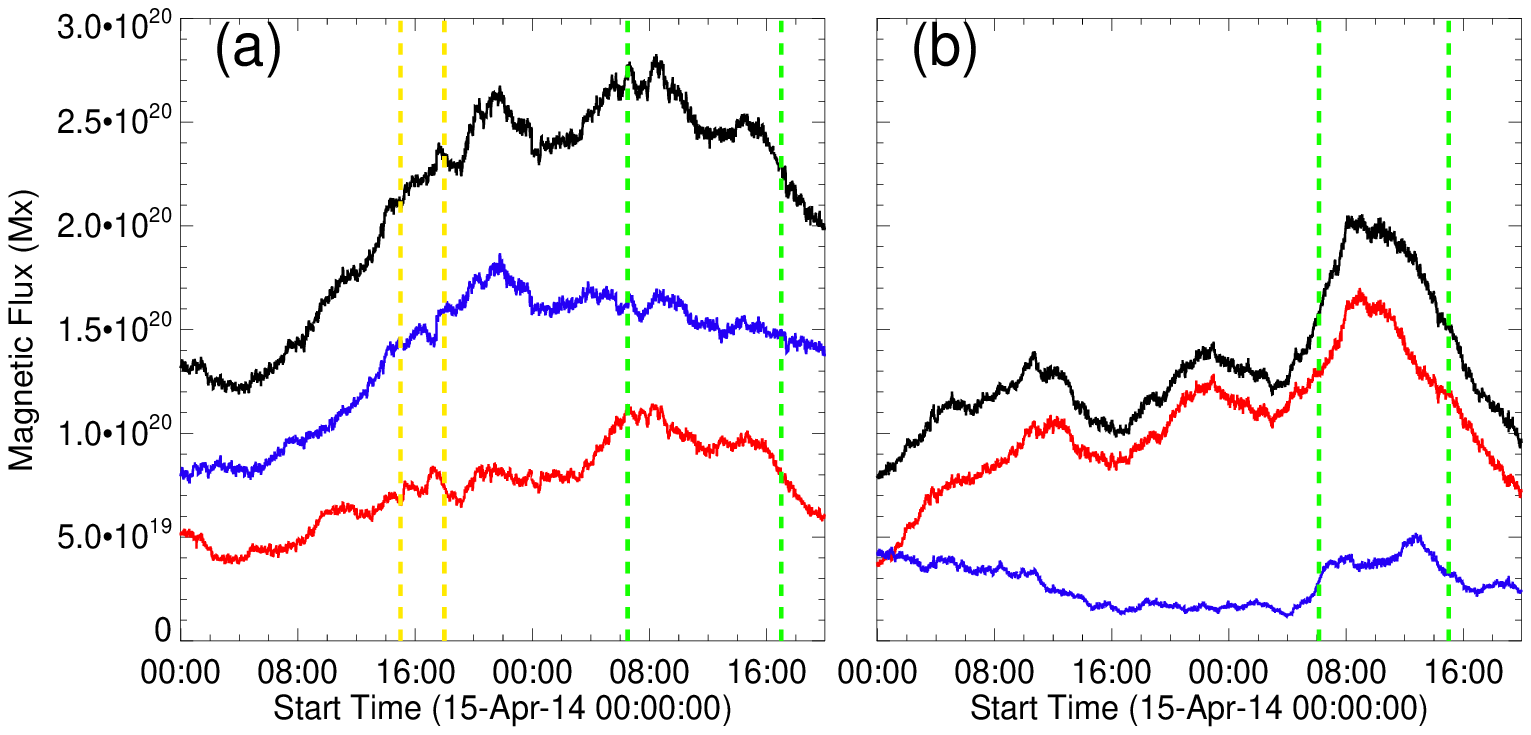}
\vspace*{-0.1cm}
\caption{Magnetic flux as a function of time calculated over the red box of
 Figure \ref{fig:mag} for the site 1 (a) and site 2 (b) location. Black, red and blue curves are for
total, positive and negative magnetic flux respectively. 
Jets on 15 April and 16 April occur in between the yellow and green dashed 
lines respectively.}
\label{fig:mag_evo}
\end{figure}

The zoomed view of magnetic flux evolution at site 1 is shown in Figure
\ref{fig:mag1}. In the
figure, we have noticed several patches of emerging flux of positive and negative polarities.
In addition to this emerging flux, we have interestingly found that the 
negative polarity N2 and the positive polarity p came closer and cancelled
each other. The jet's cancelling location is represented by the red arrow.
 To examine the flux cancellation at the jet
site 1, we made  a  time-slice diagram along the slit
shown in Figure \ref{fig:mag1} as the yellow dashed line.
The result  is presented in Figure \ref{fig:mag_slice}.
The positive and negative flux approached each other and cancelled afterwards.
We have drawn the start and end time of the jets from site 1 and this is
shown in the figure by vertical red dashed lines. We noticed that the jet
activity from  site 1 was between this flux cancellation site.
Further, we did quantitative measurements in the box at jet site 1 drawn in Figure \ref{fig:mag} (top, middle panel).
The positive, negative and total signed flux variation as a function of time
 calculated over the box is shown in Figure \ref{fig:mag_evo}a.
On 15 April the flux is constantly emerging even with some cancellation around 16:00 UT, 
on 16 April the flux decreases due to cancellation.
For site 2, the enlarged version of magnetic field evolution is shown in
Figure \ref{fig:mag2}. The emergence of  different  positive
and negative patches is shown by green and cyan arrows respectively.
 Around the site 2 location, we have observed  positive 
polarities surrounded by a kind of supergranule cell.
 Inside this positive
polarity supergranule, small bipoles with mainly negative polarities are 
continuously emerging. 
For the quantitative evolution of the positive, negative and total magnetic flux at 
the site 2 location, we have also calculated the magnetic flux as a function of time inside the
red box (top, middle) of Figure \ref{fig:mag}. The variation of magnetic flux with time
 is shown in Figure \ref{fig:mag_evo}b. We have noticed that the emergence of magnetic field
is continuously followed by the cancellation. At site 2, all the four jets are present on 
16 April and we have drawn the onset and end time of these jets. The jet duration over site 2 is
denoted by the green dashed lines in  Figure \ref{fig:mag_evo}b. We found that the jets from site 2
were lying in between the emergence and the cancellation site.


In summary, we have noticed the continuous magnetic flux 
 emergence followed by cancellation at site 1 during the time of the jet on 15 April,
 On 16 April  flux emergence and cancellation are recurrent at both sites.

\begin{table}[t]\tiny
\caption{Different physical parameters derived from SDO/AIA data for 
jets on 15--16 April 2014 at  site 1 [-220$^{''}$, -215$^{''}$] 
and site 2 [-205$^{''}$, -215$^{''}$] for different jets. The jets 
J${_1}$ -- J${_7}$ are from site 1 and
 J\ensuremath{'}${_3}$ -- 
J\ensuremath{'}${_6}$ are from  site 2. The speeds without and within parentheses are 
derived from the time--slice technique and the leading edge method respectively (see column 5).}

\label{recurrent}

\centering

\begin{tabular}{ccccccc}
\hline
Jet&~~Start/& Speed at different& Height& Width& Lifetime \\
number&end  & wavelengths ($\lambda$) in km s$^{-1}$ & (Mm) & (Mm) & (min) \\
          &(UT) &304\AA~~~211\AA~~~193\AA~~~171\AA~~~131\AA~~~~94\AA &  &\\
\hline
J${_1}$ &~~14:55/ & 205~~~~ 295~~~~~ 257~~~~~ 249~~~~~ 268~~~~~ 206& 145 &4.4 &15 \\
        &15:10 &~~~~~~~~~~~~~(252)~~~~~~~~~~~~~~~~~~~~~~~~~&    &    &  \\
\hline
J${_2}$ &~~18:01/ & 234~~~~ 177~~~~~ 199~~~~~ 232~~~~~ 221~~~~~ 235& 124 &5.2 &08 \\
        &18:09 &~~~~~~~~~~~~~(196)~~~~~~~~~~~~~~~~~~~~~~~~~&    &    &  \\
\hline
J${_3}$ &~~06:33/ & 137~~~~ 140~~~~~ 132~~~~~ 153~~~~~ 148~~~~~ 163& 202 &4.3 &23 \\
        &06:56 &~~~~~~~~~~~~~(126)~~~~~~~~~~~~~~~~~~~~~~~~~&    &    &  \\
\hline
J\ensuremath{'}${_3}$ &~~06:10/ & 113~~~~ 121~~~~~ 109~~~~~ 110~~~~~ 105~~~~~ 100& 108 &3.5 &15 \\
        &06:34 &~~~~~~~~~~~~~(100)~~~~~~~~~~~~~~~~~~~~~~~~~&    &    &  \\
\hline
J${_4}$ &~~07:13/ &136~~~~ 139~~~~~ 147~~~~~ 133~~~~~ 164~~~~~ 138& 95 &2.6 &11 \\
        &07:23 &~~~~~~~~~~~~~(147)~~~~~~~~~~~~~~~~~~~~~~~~~&    &    &  \\
\hline
J\ensuremath{'}${_4}$ &~~07:12/ & 305~~~~ 295~~~~~ 325~~~~~ 300~~~~~ 296~~~~~ 303& 116 &4.5 &06\\
        &07:18 &~~~~~~~~~~~~~(322)~~~~~~~~~~~~~~~~~~~~~~~~~&    &    &  \\
\hline
J${_5}$  &~~10:36/ & 183~~~~ 202~~~~~ 174~~~~~ 185~~~~~ 184~~~~~ 182& 217 &6.0 &24\\
        &10:50 &~~~~~~~~~~~~~(174)~~~~~~~~~~~~~~~~~~~~~~~~~&    &    &  \\
\hline
J\ensuremath{'}${_5}$ &~~10:33/ & 275~~~~ 343~~~~~ 364~~~~~ 291~~~~~ 316~~~~~ 241&87  &3.6 &10 \\
        &10:43 &~~~~~~~~~~~~~(357)~~~~~~~~~~~~~~~~~~~~~~~~~&    &    &  \\
\hline
J${_6}$ &~~14:41/ & 197~~~~ 177~~~~~ 192~~~~~ 156~~~~~ 170~~~~~ 171&94  &3.0 &15 \\
        &14:55 &~~~~~~~~~~~~~(187)~~~~~~~~~~~~~~~~~~~~~~~~~&    &    &  \\
\hline
J\ensuremath{'}${_6}$ &~~14:47/ & 343~~~~ 326~~~~~ 340~~~~~ 323~~~~~ 307~~~~~ 304& 152 &4.0 &16 \\
        &14:59 &~~~~~~~~~~~~~(335)~~~~~~~~~~~~~~~~~~~~~~~~~&    &    &  \\
\hline
J${_7}$ &~~16:59/ & 183~~~~ 174~~~~~ 154~~~~~ 187~~~~~ 184~~~~~ 217& 145 & 5.1 &14\\
        &17:13 &~~~~~~~~~~~~~(154)~~~~~~~~~~~~~~~~~~~~~~~~~&    &    &  \\
\hline
\end{tabular}
\end{table}

\section{Magnetic Topology}
 \label{mag_top}

\begin{figure}[h]
\includegraphics[width=1.00\textwidth,clip=]{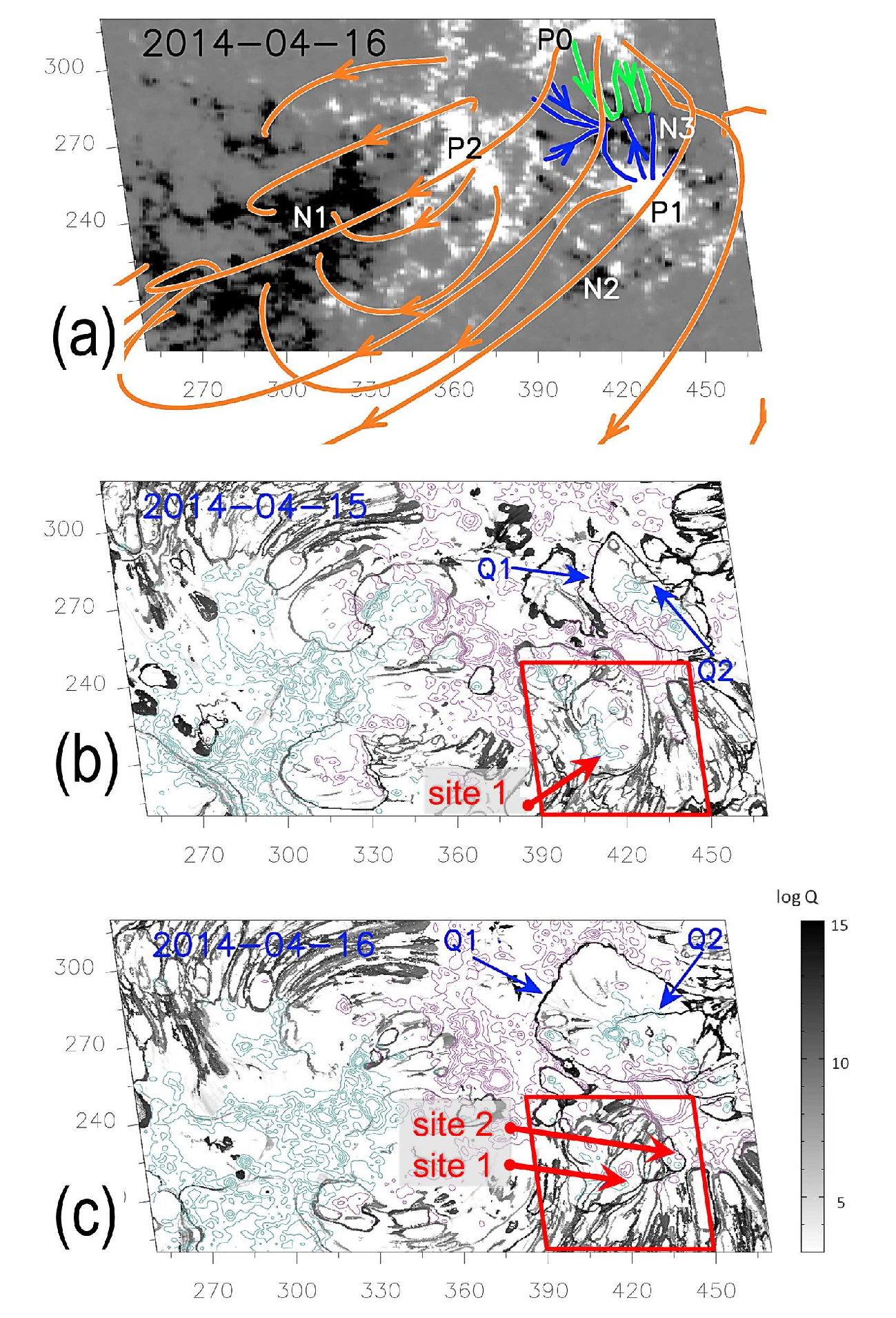}
\vspace*{-0.1cm}
\caption{Magnetic field  (line of sight component) of AR 12035 for
 16 April 2014. Panel (a)  shows the two leading positive polarities 
P1 and P2 and the following negative polarity N1. 
The magnetic field is  overlaid by magnetic field lines obtained 
from a potential extrapolation. Panels (b)  and (c):  Contours of the magnetic
 field  overlaid by the  footprints of the QSL  at {\em z}=0.4 Mm above the photosphere, 
respectively for 15 April (b) and 16 April (c). 
Q1 and  Q2 indicate the  QSLs related to the flares, 
The  magenta/cyan contours are drawn for
levels of the magnetic field  equal to $\pm 100, \pm 300, 
\pm 500, \pm 700, \pm 900$ Gauss. The red boxes indicate the region of the jets, site 1 on 15 April and both sites on 16 April.
(adapted from \inlinecite{Zuccarello17}).}
\label{fig:top}
\end{figure}

The  longitudinal magnetic field  maps observed by HMI show a strong 
complexity of the polarity pattern and a fast evolution 
(Figures \ref{fig:mag}, \ref{fig:mag1}, \ref{fig:mag2}) in the two
 sites where the two series of jets are initiated. Looking carefully at
 the AIA movies we have detected a shift of the jets  from site 2
to site 1 and never from site 1 to site 2.  The shift is not always fully accomplished  or fully observed. The best case is  jet J\ensuremath{'}${_5}$ from site 2  which completely 
merged with  J${_5}$ from site 1.
It is important to understand why there is a 
slippage of the magnetic field lines. Slippage reconnection has been 
observed in many flares \citep {Priest1992, Berlicki2004, 
Aulanier2005, Dudik2012}.
Commonly it is due to the slippage of magnetic field lines  anchored along 
QSL structures. The slippage occurs when and 
where the squashing degree is high enough along the QSL to force the 
reconnection. \inlinecite {Demoulin1996} and \inlinecite{Demoulin1998} has shown, 
theoretically and 
observationally, how it can be produced. \citet{Dalmasse2015}  demonstrated that the QSLs
 are robust structures and can be computed in potential configurations. 
Qualitatively the results are very good with this approach
and there is a relatively good fit between the location of QSL footprints
 with the observed flare ribbons \citep{Aulanier2005}. However when the
 magnetic configuration is  too complex, the QSL footprints do not fit 
perfectly and a one to one comparison is  increasingly difficult with smaller and smaller scale polarities
\citep {Dalmasse2015}.   
The benefit of using linear force--free field (LFFF)  extrapolation can be weak since QSLs are
 robust to parameter changes.  In the present case, LFFF extrapolation would perhaps help 
to follow the path of the jets which shows some curvature at their  bases  
(mentioned as a sigmoidal shape in the text of Section 2.1 ) but at the expense of the geometry of loops at large heights.
However, the magnetic field strength  is really fragmented in the jet regions. 
Pre-processing the data would smear  electric currents in their moving weak  polarities,  so it   will
  not help to derive better QSLs.
Therefore to investigate the magnetic topology of the jet producing regions, 
we use the same
 potential magnetic field extrapolation of AR 12035 calculated in \inlinecite {Zuccarello17}. 
This method is based on the fast Fourier transform method of \inlinecite {Alissandrakis1981} and
 the extrapolation is performed by using a large field-of-view that includes at its center the AR 12035. 
This allows us to identify the key topological structures of the active region (see Figure~\ref{fig:top} left panels).
Maps of the squashing degree {\em Q} at the photospheric plane were calculated using the topology tracing code 
topotr (see  \inlinecite {Priest1995}, \inlinecite{Demoulin1996}, \inlinecite{Pariat2012} for more details).
The locations of the largest values of {\em Q} define the footprints of the QSLs 
\citep{Demoulin1996,Aulanier2005} and they correspond to regions where electric current layers can easily develop.

In \inlinecite{Zuccarello17} we study the behavior of eruptive and failed eruptions
 occurring in the north-west part of the AR. We have found two QSLs: the first one (Q1 in Figure~\ref{fig:top}b and c) 
encircling the positive polarity P1 and separating the magnetic flux system from the external field and a 
second one (Q2 in Figure~\ref{fig:top}b and \ref{fig:top}c) highlighting
the spine of a high-altitude coronal null-point similarly to what is seen in \inlinecite{Masson2009}. 
The flares occurred mainly at the north-west edge of the large QSL (Q1).

Since the fan-like QSL (Q1) encircling the flare region was separated from the complex QSL system 
around the jet producing region (at the south of P1), we have argued that the jet activity and the 
flares were not really linked to each other, even if their timings seem to be related \citep{Zuccarello17}.

\begin{figure}[h]
\includegraphics[width=1.00\textwidth,clip=]{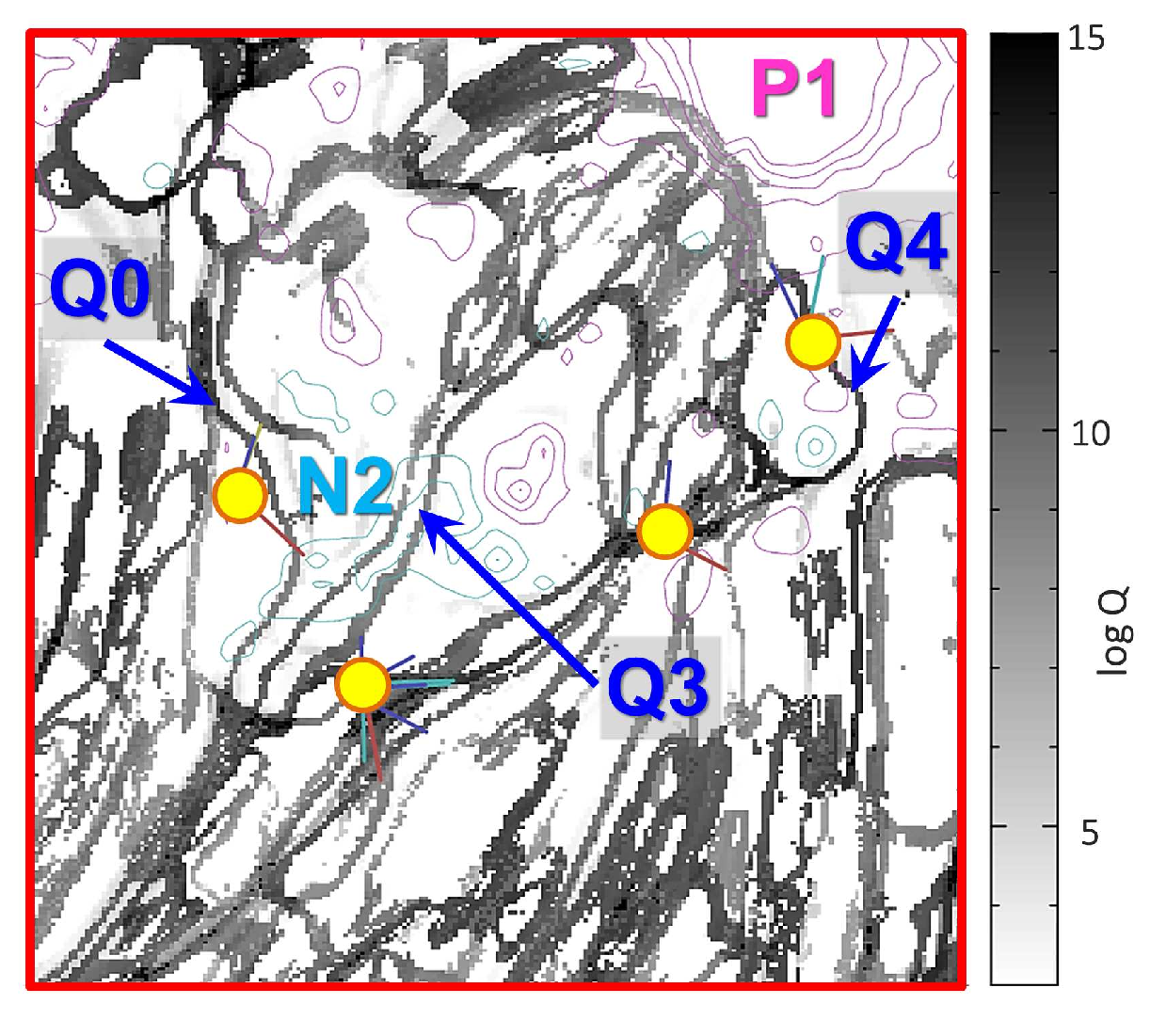}
\vspace*{-0.1cm}
\caption{Zoom of the jet region 
 inside  the red box  of 16 April (see  panel (c) in Figure \ref{fig:top}).
 Q1 and  Q2 indicate the  QSLs related to the flares, 
Q3 and Q4 to the site 1 and  site 2  of the jets
respectively.  The vertical color bar indicates the value of the
squashing degree {\em Q}. The yellow circles  indicate the positions 
of four  null points. }
\label{fig:null}
\end{figure}

On 15 April the jets are initiated in site 1 and we find  a well defined QSL surrounding the region of the jet.
 On 16 April the configuration is much more complicated with 
many QSLs  which are in the site region  of both jets.  The zoom of the {\em Q} map of 16 April
 around the jet producing region (red box of Figure~\ref{fig:top}c) is shown in Figure~\ref{fig:null}. 
We find that the two sites of the jets, site 1 and site 2, are respectively inside the QSLs Q3 and Q4, 
and that both were embedded in a larger QSL, labeled as Q0 in the figure. 
We also identified several quasi photospheric null points, that are indicated by yellow circles. 
The QSL map is very complex, and difficult to analyze and compare in detail with the observations due to 
the small scale of the events and fast motion of the small polarities, both the moving p polarity in 
site 1 and the emergence of small bipoles in site 2.
Since it looks quite possible that the 2 QSLs (Q3 and Q4) intersect or touch each other, 
we conjecture that field line foot points could move from site 2 and site 1 by a sequence of 
reconnections across QSLs as in \inlinecite{Dalmasse2015}. 
This could produce the transfer/or tendency of movement of jets from site 2 to site 1, as we have
 observed for jets J$_5$ and J\ensuremath{'}${_5}$ for example (Figure~\ref{fig:jet5}). In the case of the
 other jets from site 2, they also show a tendency of slippage towards site 1.
 On 15 April the topology of  QSLs in the jet region is more simple.
 That could explain perhaps why there are no jets from site 2.

\section{Discussion and Conclusions}
     \label{result}

In this study, we have analyzed eleven solar jets originating from two 
different sites in the same active region
 (NOAA AR 12035) during 15--16 April 2014 using the 
SDO/AIA, HMI, and ARIES H$\alpha$ data. 
The magnetic topology of the active region was discussed using a potential 
magnetic field extrapolation. 
The extrapolation was done using HMI photospheric magnetic 
field as the boundary condition.
The main conclusions of our study are as follows:

\begin{itemize}
\item 
We found two sites for the different jet's activity. 
On 15 April the jets originated from site 1 and we measure  a large increase of  
emerging flux and small cancellation. On 16 April  site 1 
and site 2 are  associated with continuous  emerging magnetic flux followed by  cancellation at the jet time. 

\item 
The kinematics of jets at different  EUV wavebands revealed that the speeds, 
widths, heights and lifetimes of jets
are slightly different at different wavelengths. This can be interpreted as
 the multi--temperature and multi-- velocity structure of
solar jets.

\item
Most of the jets showed clockwise rotation, which
 indicates untwisting. As a result of this 
untwisting, the twist/helicity was injected 
in the upper solar atmosphere.

\item
We observed the slippage of jets at site 2 namely
 J\ensuremath{'}${_3}$--J\ensuremath{'}${_6}$ towards the eastern site (site 1) and never the reverse movement.
Along with the movement of jets towards site 1, we found, in the case of 
jet J\ensuremath{'}${_5}$ that a part
 detached from it and moved towards the site 1 location and
finally merge into jet J${_5}$.

\item
On 16 April both  jet sites are associated with 
the QSLs. The possible intersection of the two
 QSLs encircling each site could explain the slip 
reconnection occurring along the
 QSLs  which favor the translation of jets from
 site 2 to  site 1.

\end{itemize}

The clockwise rotations (right-to-left) in some of the jets 
indicate the untwisting of the jets.
The untwisting jets eject the helicity in the
 higher solar atmosphere \citep{Pariat15}.
The injected helicity in the jets may
be  part of the global emergence of  twisted magnetic fields.
During the rotation like in the case of jet J${_1}$, we observed 
the rotating  jet material contains bright as
 well as dark material.
This result is consistent with simulations done by \inlinecite{Fang14}.
In their simulation, they found the simulated jet consists 
of untwisted field lines, with a mixture of cold and hot plasma.

The kinematics of the jets indicates that different jets have not only
different speeds but their speed also varies with different wavelengths.
This can be interpreted as  multi-temperature and
multi-velocity structures in the solar jets.
 Our calculated values of the speeds, widths
and lifetimes are consistent with earlier reported values
in the literature \citep{Shimojo96,Shimojo00,Schmieder13,Chandra15}.
We have also observed that the average lifetime is longer in 304 \AA\
than in shorter wavelength observations,
which suggests that the cooler component of jets have a longer  lifetime
in comparison to the hotter component.
This supports the study of \inlinecite{Nistico09}.
 \inlinecite{Nistico09} compared the lifetime between 171 \AA\ and 
 304 \AA\ and found that the lifetime is longer for the longer wavelength. 
For all the studied jets, except for 
J${_5}$ and J\ensuremath{'}${_5}$, the jet peak time is simultaneous at all wavelengths.
 In the case of  J${_5}$ and
 J\ensuremath{'}${_5}$, the peaks at longer wavelengths are earlier than
at the shorter EUV wavelengths. The time delay between longer and shorter EUV wavelengths
 in the case of jet J${_5}$ and J\ensuremath{'}${_5}$ can be interpreted as
during the reconnection, there could be a different heating time for
different threads.

Thanks to the SDO high spatial and temporal resolution,
  we could examine the
dynamics of these jets in a more precise way.
As mentioned in Section \ref{obs1}, the detached part of J\ensuremath{'}${_5}$
moved  towards J${_5}$ and finally merged with J${_5}$.
The motion of  the broken part of J\ensuremath{'}${_5}$
towards the east can be interpreted
as the expansion of the reconnection region with time \citep{Raouafi16}.
As suggested by \inlinecite{Sterling15}, the coronal jets
may be due to the eruption of mini filaments and they predicted
that the spire of the jets moves away from the jet base bright point.
Our observation of the motion of the broken part of  J\ensuremath{'}${_5}$
is away from the jet base. This supports the findings of \inlinecite{Savcheva09} 
and the interpretation proposed by \inlinecite{Sterling15}.

We have observed the flux emergence followed by flux cancellation at site 1 during 15 April 2014.
Moreover, on 16 April 2014 flux emergence and cancellation are recurrent in both jet sites.
The observation of cool and hot material in our study
supports the hypothesis of small filament eruption and 
a universal mechanism for eruptions at different scales \citep{Sterling15,Wyper2017}.
\newpage

\newpage

\newpage

\newpage

\newpage

\newpage

\newpage

\newpage

\newpage

\newpage

\begin{acks}
The authors are thankful to the referee for the
 constructive comments and suggestions
which improved the manuscript significantly.
We acknowledge the SDO/AIA and HMI open data policy. RJ 
acknowledges the Department of 
Science and Technology (DST), Government of India 
for an INSPIRE fellowship. This work was initiated during 
the one month stay of RC at the Observatoire de Paris, LESIA, Meudon, France.
RC also acknowledges the support from SERB--DST project no. SERB/F/7455/ 2017-17.

\noindent {\bf Disclosure of Potential Conflicts of Interest} The authors declare that they have no conflicts of interest.  
\end{acks}

\end{article}
\end{document}